\documentclass[apl,aps,prb,preprint,groupedaddress,preprintnumbers,showpacs]{revtex4}

\usepackage[T1]{fontenc}
\usepackage{graphicx}
\usepackage{dcolumn}

\usepackage{color,graphics,graphicx,fancybox,fancyhdr}
\usepackage{latexsym}
\usepackage{amsmath}
\usepackage{amssymb}

\begin{document}
\bibliographystyle{apsrev}

\graphicspath{../Latex} \DeclareGraphicsExtensions{.eps,.ps}

\title{Intrinsic nonlinearity probed by intermodulation distortion microwave measurements on high quality MgB$_{2}$ thin films}

\author{G. Cifariello, M. Aurino, E. Di Gennaro, G. Lamura$^{a}$ and A. Andreone}

\affiliation{CNR-INFM Coherentia and Department of Physics,
University of Naples Federico II, I-80125, Napoli, Italy.}

\author{P. Orgiani$^{b}$, X. X. Xi}

\affiliation{Department of Physics, The Pennsylvania State
University, University Park, PA 16802, U.S.A.}

\author{J.-C. Villégier}

\affiliation{CEA-Grenoble SPSMS/LCP, F38054 Grenoble Cedex 9,
France.}

\date{\today}
\begin{abstract}
The two tone intermodulation distortion arising in Mg$B_{2}$ thin
films synthesized by hybrid physical-chemical vapour deposition is
studied in order to probe the influence of the two bands on the
nonlinear response of this superconductor. The measurements are
carried out by using a dielectrically loaded copper cavity
operating at 7 GHz. Microwave data on samples having critical
temperatures above 41 K, very low resistivity values, and residual
resistivity ratio larger than 10, are shown. The dependence of the
nonlinear surface losses and of the third order intermodulation
products on the power feeding the cavity and on the temperature is
analyzed. At low power, the signal arising from distortion versus
temperature shows the intrinsic $s$-wave behavior expected for this
compound. Data are compared with measurements performed on Nb and
YBa$_{2}$Cu$_{3}$O$_{7 - \delta}$ thin films using the same
technique.

\end{abstract}

\pacs{74.70.Ad,74.25.Nf}

\maketitle

A major factor limiting the performance of passive superconducting
devices is the nonlinear response to an applied r.f. field,
including the generation of harmonics and IMD products. The nonlinear
response of a superconductor has two possible and different
origins: extrinsic (like the presence of grains and grain
boundaries \cite{FINDI}) and intrinsic (the nonlinear Meissner
effect, NLME \cite{YIP,XU}). The extrinsic nonlinear properties
of a superconductor are generally associated to the presence of
weak links: the Josephson coupling between grains lowers the first
penetration field and increases the level of harmonics and
intermodulation distortion (IMD) generation in both $d$-
\cite{FINDI} and $s$-wave \cite{LAMURAIII} superconductors. This
effect can be safely excluded in epitaxial thin films, where the
intrinsic origin of nonlinearity is the only and unavoidable
cause degrading the performances of whichever superconducting
electronic device.

The intrinsic nonlinear response arises from the backflow of
excited quasiparticles at finite temperatures well deep in the
Meissner state. Some years ago, Dahm and Scalapino
\cite{DAHMI,DAHMII} showed that at low temperatures and small
fields the presence of quasiparticles changes the nonlinear
response of a superconductor in a way that is remarkably different
depending on the pairing symmetry. In the case of magnesium
diboride, they also predicted what would be the influence of the
two bands on the IMD behavior \cite{DAHMIII}. More recently, this
issue has been analyzed thoroughly by considering also the role of
anisotropy and strong coupling effects\cite{CARBOTTEI}. Such
theoretical studies give therefore the possibility to probe the
gap function by looking at the non linear response of a
superconductor in the microwave region. This has been done quite
recently by using microstrip resonant techniques on patterned
niobium and YBa$_{2}$Cu$_{3}$O$_{7 - \delta}$ (YBCO) thin films
\cite{BENZ,OATESIII}. In this paper we present an experimental
study of the NLME in MgB$_{2}$, using IMD measurements carried out
in a dielectically loaded cavity on high quality thin films
synthesized by hybrid physical-chemical vapour deposition (HPCVD).
We compare the results with data taken on YBCO and Nb films.
Contrary to techniques based on transmission-line resonators, this
method probes the microwave properties of \textit{unpatterned}
samples, therefore minimizing the enhancement of current density
at the edges, that in principle might rise to extrinsic effects in
the electrodynamic response.

MgB$_{2}$ films were grown on (0001)4H-SiC. The substrate choice
is due to its very small (-0.45\%) lattice mismatch with magnesium
diboride. A detailed description of the HPCVD technique has been
reported elsewhere \cite{XiI,XiII}. X ray diffraction shows that
the films grow epitaxially, with the c-axis oriented normal to the
substrate and the a-axis parallel to the a-axis of SiC
\cite{XiI,XiIII}. These samples show values of $T_{C}$ higher than
those reported for bulk materials, because of coherent strain
induced by the epitaxial growth \cite{XiIV}. Moreover they are
very pure, as indicated by the low residual resistivity (minimum
value of $0.26 \mu \Omega·cm$).

To investigate the power dependence of the microwave properties of
the MgB$_{2}$ films, we performed in the same system configuration
the measurement of the surface losses and of the two-tone
intermodulation as a function of the input power at different
temperatures below T$_{c}$. We used an open-ended dielectric
single-crystal sapphire puck resonator, similar to the one used in
ref.\cite{LAMURAIII}, excited with a transverse electric mode
(TE$_{011}$) and operating at the resonant frequency of 7 GHz. The
resonator enclosure is made of Oxygen Free High Conductivity
(OFHC) Copper, as well as the sample holder, placed in the center
of the cavity in close proximity with the dielectric crystal.  The
sapphire puck (8 mm height and 16 mm diameter) is separated from
the copper wall by a thin sapphire spacer (6 mm height and 2.5 mm
diameter). By using a micrometer screw, the puck-to-sample
distance can be changed, depending on the material under test, in
order to get the maximum sensitivity. The cavity is taken under
vacuum using a copper can, which includes a double layer
$\mu$-metal shield, and inserted in a liquid helium cryostat. For
the measurement of IMD third order products, two closely spaced
tones with equal amplitudes at frequencies $f_{1}$ and $f_{2}$ are
generated by two phase-locked CW synthesizers. The signals,
symmetrically separated around the center frequency of the cavity
by an amount $\Delta f$, are combined and applied to the resonant
structure. All IMD data presented here are taken with $\Delta f$ =
10 kHz, whereas the 3 dB resonance bandwidth is at least a factor
10 larger at all temperatures below $T_{C}$. The output signals
coming from the cavity (the two main tones at $f_{1}$ and $f_{2}$
and the two third-order IMDs at 2$f_{1}$ - $f_{2}$ and 2$f_{2}$ -
$f_{1}$) are detected using a Spectrum Analyzer. No amplifier is
used to avoid unwanted nonlinearities. The small puck-to sample
distance (less than 1 mm) ensures a very good signal-to-noise
ratio even at very low input power level. The first detectable
signal from the third order intermodulation products starts at
-140 dBm or less.

We measured four MgB$_{2}$ films, whose deposition conditions were
slightly different, resulting in residual resistivity ratio (RRR)
values always greater than 10, and thickness values between 150
and 400 nm. All the samples under test have the same critical
temperature ($\sim$ 41 K) and a similar behavior as a function of
power. For the sake of clarity we will show the experimental
results obtained on one sample only, which is the most carefully
characterized. This film is 150 nm thick, with an onset $T_{C}$=
41.4 K, $\Delta T_{C}$= 0.1 K. The surface resistance measured at
low power levels and at 4 K is below our instrument sensitivity
($50 \mu\Omega$).

To probe the intrinsic nonlinear properties of this
superconductor, we carefully checked under which conditions the
Meissner regime applies, i.e. the applied field is lower than the
first penetration field $H^{*}$. To this aim, we have measured the
variation of the surface resistance as a function of the applied
microwave field at a fixed temperature from 4 K up to the critical
temperature. The average microwave field feeding the cavity has
been calculated from the circulating power~\cite{LAMURAIII}. In
fig.1 $\Delta R_{S}$ data versus $H_{RF}^{2}$ at 4, 7, 18 and 30 K
respectively are plotted. The quadratic behavior predicted for
$\Delta R_{S}(H_{RF})$ in the Meissner state is clearly observed,
even if limited to a very small field range at all temperatures.
The arrows in the figure indicate the first penetration field
$H^{*}$, marking the transition from a quadratic dependence to a
linear or sublinear one. The difference in the behavior is
confirmed also by measurements of the parameter r = $\Delta
R_{S}(H_{RF})$/$\Delta X_{S}(H_{RF})$~\cite{GOLO}. For $H_{RF}$ <
$H^{*}$, r << 1 (in the range $10^{-2}$), as it has to be since in
the Meissner regime the nonlinear inductive reactance dominates
the nonlinear response. For $H_{RF}$ > $H^{*}$, r $\sim$ 1, with
losses from normal regions driving the electrodynamic behavior of
the samples. This has been previously ascribed as due to vortex
penetration in grain boundaries~\cite{LAMURAIII}. The $H^{*}$
values correspond to circulating powers of 23, 20, 18 and 16 dBm
respectively for the temperatures plotted in fig. 1. Below these
power values, the IMD response versus the power circulating inside
the cavity ($P_{circ}$) show the expected intrinsic behavior
(slope 3) at all temperatures under test. Only at 30 K data start
to depart from a cubic dependence at the highest power levels. In
low quality magnesium diboride films grown by dc sputtering,
extrinsic effects dominate and the dependence of the IMD products
on the r.f. power clearly follows a quadratic law (slope
2)~\cite{LAMURAIII}.

In fig. 2 the power $P_{IMD}$ arising from third order products in
the MgB$_{2}$ film at 7 K is displayed as a function of
$P_{circ}$. The results are compared with similar measurements
performed at the same reduced temperature (t = T/T$_{C}$ $\sim$
0.17) on a high quality epitaxial Nb film (200 nm thick) deposited
by dc magnetron sputtering~\cite{VILLEGIER} and two highly
oriented YBCO films (700 nm thick) grown by thermal evaporation at
Theva GmbH. Data are normalized using the factor
$P_{N}=10^{-\frac{IL}{20}}Q_{L}$ (IL are the insertion losses
expressed in dB and $Q_{L}$ is the unloaded quality factor) since
this allows to remove the dependence of the experimental results
on external conditions~\cite{OATESIII}. The details of the
normalization procedure and of these other experiments will be
presented elsewhere.

Following Dahm and Scalapino, the nonlinear response of a
superconductor can be described looking at the normalized IMD
power because of the relation:
\begin{eqnarray}\label{1}
\frac{P_{IMD}}{P_{N}}= K \cdot b^{2}(t) \cdot P_{circ}^{3}.
\end{eqnarray}

The quasiparticle backflow contribution of the NLME lies in the
nonlinear parameter $b(t)$, that displays a different temperature
behavior depending on the gap function symmetry. $K$ is a factor
related exclusively to the parameters of the material under
test\cite{DAHMII,CARBOTTEI}. In particular, it strongly depends on
the zero temperature London penetration depth $\lambda(0)$,
possibly explaining the large difference in the absolute values
observed in fig. 2. A similar scattering of data has been
previously observed in YBCO films \cite{OATESIV}. Nevertheless, a
major feature in the figure is that MgB$_{2}$, as confirmed also
by measurements performed on the other samples, shows a higher
level of intrinsic nonlinearities at low temperature and low
circulating power in comparison with both YBCO and Nb.

If we divide eq. 1 by $P_{Circ}^{3}$, the result becomes
independent of the energy level of the e.m. excitation feeding the
cavity. The experimentally accessible quantity will depend
therefore on $b(t)$ and $K$ only. If we re-scale our experimental
data on the theoretical
curves\cite{DAHMI,DAHMII,DAHMIII,CARBOTTEI}, we indirectly assume
that $K$ has the same $\lambda(0)$-dependence of the theoretical
model. In fig. 3 a comparison of the measurements reported here is
shown, plotting the nonlinear coefficient $b^{2}$ evaluated at
constant circulating powers as a function of temperature for
MgB$_{2}$, Nb and YBCO. In the case of the magnesium diboride
sample, the dependence is extracted at $P_{Circ}$ = 0 dBm. All
experimental data well match the theoretical
models\cite{DAHMI,DAHMII,DAHMIII,CARBOTTEI} (solid line, one-band
$s$-wave; dash-dotted line, dirty two-band $s$-wave; dotted line,
$d$-wave). In the case of YBCO thin films, the nonlinear parameter
increases as $1/T^{2}$ following the prediction for the gap
function of a $d$-wave superconductor, and in agreement also with
other experiments reported in literature\cite{OATESIII}. In
MgB$_{2}$, as well as in Nb, $b^{2}(t)$ decreases monotonically
towards zero, indicating in both cases an $s$-wave behavior. The
small residual value at the lowest temperatures is an indication
of impurities and/or strong coupling effects \cite{CARBOTTEI}. We
do not observe any feature in the $b^{2}(t)$ behavior for
MgB$_{2}$ that can be associated with decoupled bands, implying
that in high quality samples there is a moderately large value of
interband coupling\cite{CARBOTTEI}. Moreover, using the analytical
formulas reported in ref.\cite{DAHMIII}, the experimental data
seems to match very well the theoretical curve assuming a finite
intraband scattering rate $\Gamma$ for both $\sigma$ and $\pi$
bands, in particular yielding a ratio
$\Gamma_{\pi}/\Gamma_{\sigma}$ close to 2.

To conclude, it is worth noting that the peculiarity of MgB$_{2}$
is that the level of nonlinearity can be decreased changing the
scattering rates in the two bands\cite{DAHMIII, CARBOTTEI}. We
believe that, playing with substitutional doping, there is still
room for improvement, in view of a possible application of
magnesium diboride for the development of new passive
superconducting devices.

\begin{acknowledgments}
The work at the Department of Physics of the Pennsylvania State
University is supported in part by NSF under grant No. DMR-0306746
and by ONR under grant No. N00014-00-1-0294. The work at
University of Naples \textit{Federico II} is supported in part by
PRIN 2004 "Two-gap superconductivity in MgB$_{2}$: the role of
disorder".
\end{acknowledgments}

$^a$Author to whom correspondence should be addressed. E-mail:
gianrico.lamura@na.infn.it.

$^b$present address: Dipartimento di Fisica "E. R. Caianiello" and
Laboratorio Regionale SuperMat INFM-Salerno, Università degli
Studi di Salerno, Baronissi (Sa) I-84081, Italy.

\newpage


\newpage

\textbf{Figure captions}

Fig. 1: Variation of the surface resistance as a function of the
applied microwave field at different temperatures. Arrows set the
limit of the quadratic regime.\\

Fig. 2: The normalized output power of the IMD products as a
function of the circulating power for the MgB$_{2}$ thin film
under test ($\triangle$); an epitaxial Nb thin film ($\Box$);
standard high quality YBCO films ($\bullet$, $\circ$); at the same
reduced temperature (t = T/T$_{C}$ $\sim$ 0.17). Continuous line
illustrates slope 3.\\

Fig. 3: The behavior of $b^{2}$ versus the reduced temperature t =
T/T$_{C}$. The dotted line represents the theoretical behavior for
the $d$-wave case, the solid line for the $s$-wave one-band case,
and the dash-dotted line for the $s$-wave two-band case with
$\Gamma_{\pi}/\Gamma_{\sigma}=2$; points represent the
experimental results for standard high quality YBCO films
($\bullet$, $\circ$); an epitaxial Nb thin film ($\Box$); the
MgB$_{2}$ thin film under test ($\triangle$) at a circulating
power of 0 dBm.\\

\newpage

\begin{figure}[h]
\includegraphics[scale=0.95]{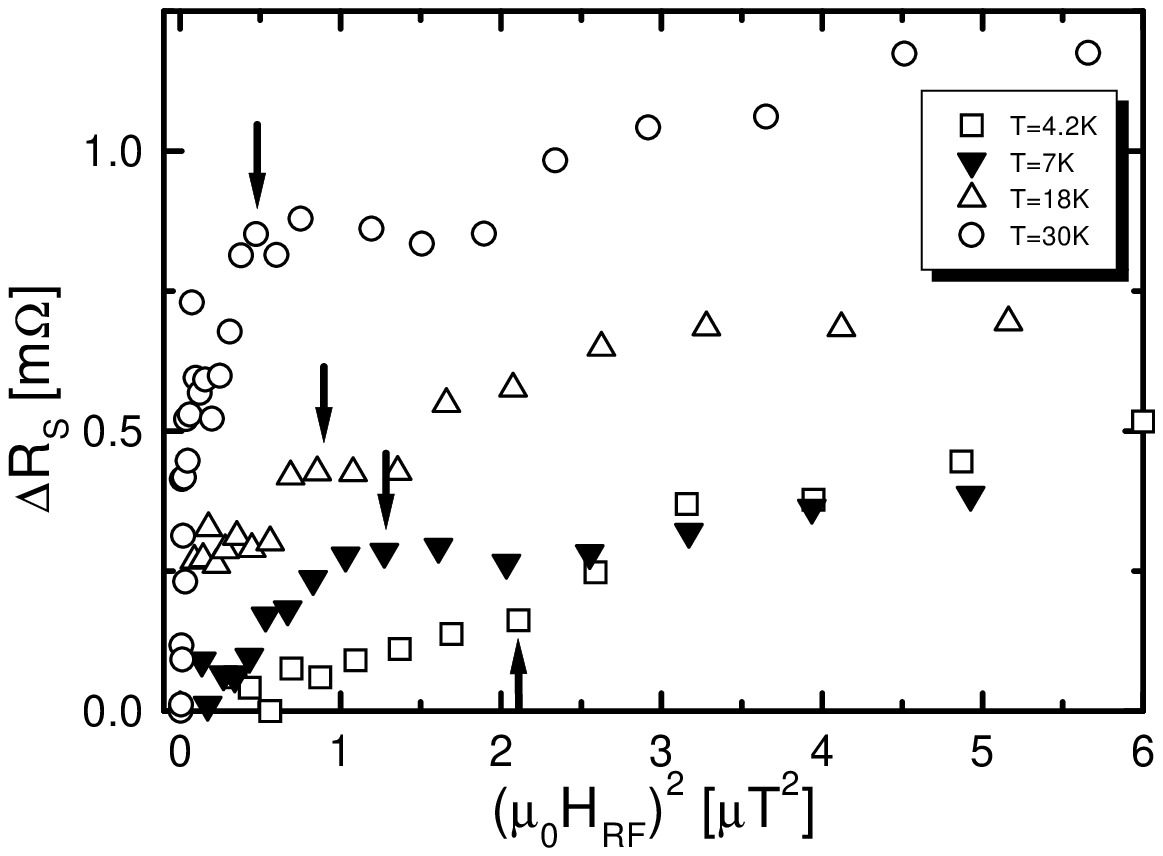}
\end{figure}

\begin{figure}[h]
\includegraphics[scale=0.95]{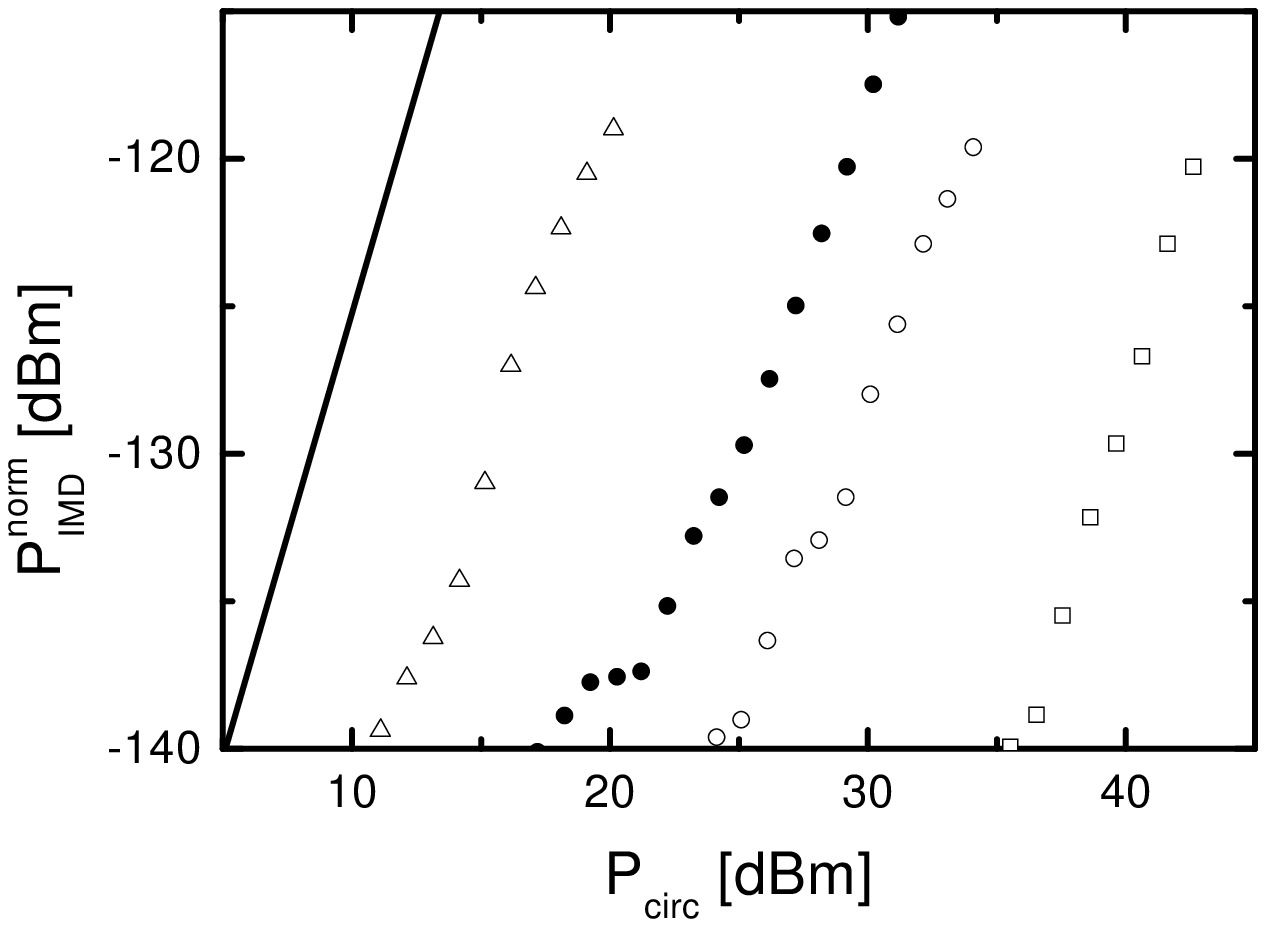}
\end{figure}

\begin{figure}[h]
\includegraphics[scale=0.95]{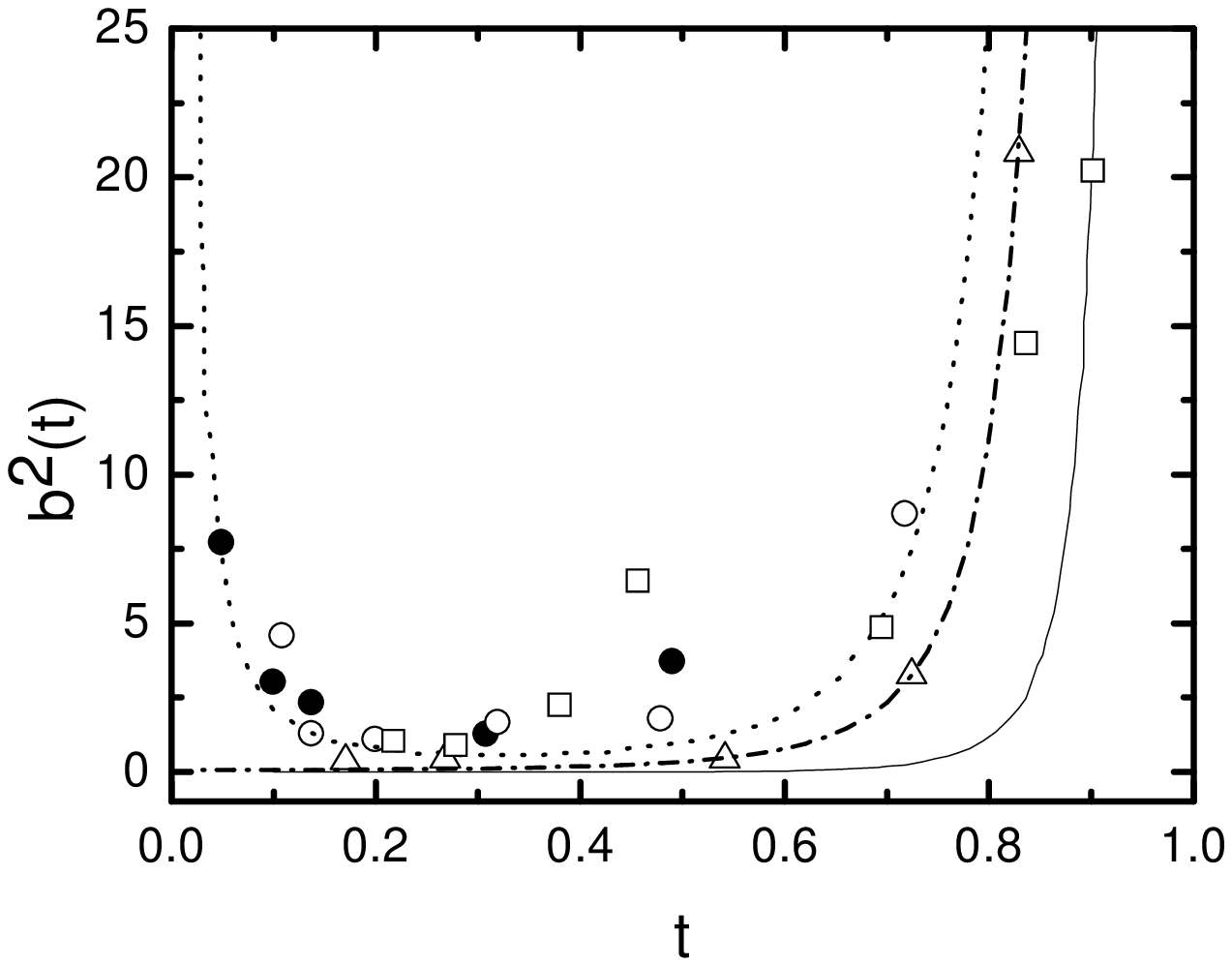}
\end{figure}

\end{document}